\documentclass[twocolumn,superscriptaddress,nofootinbib]{revtex4}
\usepackage{epsfig}
\newcommand{\ve}{\varepsilon}
\newcommand{\e}{{\rm e}}
\begin{document}

\title{
Instanton-induced Semi-hard Parton Interactions
\\and Phenomenology of High Energy Hadron Collisions
}

\author{G. W. Carter}
\affiliation{Department of Physics, University of Washington,
Seattle, WA  98195-1560}
\author{D. M. Ostrovsky}
\author{E. V. Shuryak}
\affiliation{Department of Physics and Astronomy, State University of New York,
Stony Brook, NY  11794-3800}

\date{\today}

\begin{abstract}
We phenomenologically study whether partonic collisions
responsible for the growth of hadron-hadron cross sections at high energy
can be ascribed to instanton-induced processes.
Although non-perturbative in nature, these interactions occur at the
semi-hard scale $Q\sim 1-2$ GeV, and should therefore be described 
using information from deep inelastic leptonic scattering on the 
partonic constituents in nucleons, pions, and photons.
After considering shadowing corrections in nucleon-nucleon scattering,
we fix a free instanton tail suppression parameter and determine the effective
quark-quark cross section.
The resulting contributions to $NN$, $\pi N$, $\gamma N$, and 
$\gamma\gamma$ cross sections all $increase$ with energy
differently, but in reasonable agreement with experimental data.
We then proceed to an estimate of the number of such processes 
present in high energy Au-Au collisions at RHIC, finding that 
the amount of entropy produced by instanton/sphaleron events
matches the observed amount.
\end{abstract}
\maketitle

\section{Introduction}
The theoretical description of high energy hadronic
processes began in the 1960's, when Regge-based phenomenology was developed
to describe energy dependence of interaction amplitudes.  
A special role has been played by the so-called {\em Pomeron},
the leading Regge pole with vacuum quantum numbers.
It was first believed that its intercept, $\alpha(t=0)$, was unity,
corresponding to asymptotically constant cross sections and satisfying 
the Pomeranchuck theorem, $\sigma_{pp}-\sigma_{\bar p p} \ge 0$.
However, it was discovered in the late 1970's that the cross sections grow
slowly with $s$, rendering the {\em supercritical Pomeron} with the
intercept above 1, with the precise value \cite{DL} 
\begin{equation}
\alpha(t=0)-1=\Delta\approx 0.08 \,.
\end{equation}
The discovery at  HERA of much stronger growth with energy in hard
processes, with an effective power of about $0.5$, has led 
to a proposed separation of the former ``soft'' and new ``hard'' Pomerons,
each with different parameters and different physics
\cite{two_pomerons}, the latter
presumably described by BFKL resummation of perturbative QCD
which indeed leads to a power of such magnitude \cite{BFKL}.

In this paper we will not discuss the issue of energy dependence of 
hard processes, focusing rather on the original ``soft''  Pomeron.
We put ``soft'' in quotation marks here because we do not entirely agree 
with this terminology. 
It is now clear that the Pomeron itself is a small
object, with its size represented by the slope of its trajectory,
$\alpha'(t=0) \approx 1/(4 \, {\rm GeV}^2)$.
The scale involved, $0.1$ fm, is much smaller than hadronic radii, and so
the Pomeron exchanges should in fact be treated on the level of
individual partons, appropriately defined at the 
intermediate momentum scale of 1-2 GeV.
For lack of a better standard term, we will refer to it
as the {\em semi-hard} scale.

More precisely, we will not consider the nature of the
soft Pomeron in full either. The leading Regge pole, if it exists,
is the analog of a single bound state appearing in the $t$-channel
as a result of rather different interactions\footnote{Similar to
the $J/\psi$, a definite charmonium state which appears as a result
of interplay between both perturbative and confining potentials.}.
Although the existence of such a pole is an appealing possibility,
no general principles demand that it occur in QCD.

We will follow the recent tendency of splitting the
amplitude into two parts, the {\em constant} and {\em growing} 
contributions to the cross section.
While the former part is believed to be related to color exchanges between 
partons, which lead to multiple hadron production via string 
breaking \cite{LN}, the latter is related to rarer processes resulting 
in ``prompt'' production of additional gluons, quarks, or hadrons.
Below we will also try to disentangle these two components using the
available data, and focus on the nature of the $growing$ part of 
hadronic cross sections.

The theoretical explanation of any process which takes place 
at the semi-hard scale is notoriously difficult -- which is not 
surprising, since both the pQCD and low energy 
hadronic descriptions fail at this scale.
There are basically three distinct approaches:

(i) {\em Minijet-based models} use familiar formulae from pQCD \cite{minijets}. 
They are well-tested in the domain of hard jets, but their application at
the semi-hard  scale is a drastic extrapolation. 
All of these models assume the existence of a non-perturbative
momentum cutoff, $p_{cutoff}$, in order to render pQCD results finite. 
This cutoff is left unexplained, treated as a purely phenomenological 
parameter, and all results depend greatly on its value.

(ii) {\em Instanton-based} dynamics, to be discussed below, have only
recently been applied to high-energy scattering
\cite{SZ,KKL,NSZ} and use insights obtained a decade ago in
electroweak theory \cite{weakinst}.
Particularly relevant for this work are the first two references, in which 
the growing part of the hadron-hadron cross sections is ascribed to multi-gluon
production via instantons.
 
(iii) The {\em Color Glass Condensate}, a classical Weitzecker-Williams
field of gluons carried by interacting hadrons, can be excited to produce
prompt gluons \cite{KV}.
This is another example of a weakly-coupled system involving non-perturbative
gauge field configurations.

The instanton approach, (ii), qualitatively relates the
properties of the Pomeron to other non-perturbative phenomena at low
energies \cite{NSZ} and to the static properties of the QCD vacuum
(for a review and original references see \cite{SS_98}), providing useful
constraints on parameters.
First, the ``soft'' Pomeron's compactness follows 
from the small average instanton size, $\rho\sim 1/3$ fm. 
Second, a natural explanation of the smallness of the intercept $\Delta$
(alternatively, the effective quark-quark cross section, as explained below) 
arises in that it is proportional to the ``instanton diluteness parameter'' 
of the QCD vacuum, $\kappa = n\rho^4 \sim 0.01$.
Furthermore, unlike in pQCD, at the classical level the ``odderon'' does 
not appear, since each instanton field belongs to an SU(2) 
subgroup of the SU(3) color group and therefore cannot discern between
quarks and antiquarks.

The smallness of the instanton-induced amplitudes does not 
imply an extra penalty for the production of {\em multiple} prompt
gluons and quarks.
On the contrary, such processes dominate quasielastic
(and other few-body) parton scattering. 
Instanton-induced processes furthermore lead to the creation of sphaleron-like 
gluomagnetic clusters which decay into many partons.
Thus instanton effects can be expected to overshadow perturbative
amplitudes of sufficiently high order and contribute substantially to
the prompt entropy in heavy ion collisions. 

The aim of this work is not to debate the theoretical
issues, but to try to devise a phenomenological model
capable of connecting many pieces of information about
high energy collisions. 
Therefore, in judging the approaches to semi-hard dynamics mentioned above,
we are most interested in their ability to explain the observed phenomena.
To date, there is no direct evidence which empirically supports one over
the other. 
While it is not possible to observe mini-jets, the CGC, or sphalerons directly,
one might discern between these two mechanisms by comparing their 
predictions for particle production with data.
This would require correlation analysis, which goes far beyond the 
current paper.

The goal of this paper is rather more modest.  
We analyze available hadron-hadron data and find a description 
of semi-hard interactions which involves prompt production via instantons.
We phenomenologically fix the parameter left uncertain in Ref.~\cite{NSZ},
an instanton tail cutoff, and find the effective quark-quark cross
section.
With this in hand, we compute cross sections for various hadron-hadron
scattering processes and then consider heavy ion collisions, making rough 
predictions but using no additional parameters.

\section{Shadowing in hadron-hadron collisions }

If all hadronic processes are dominated by a common Pomeron pole applied at 
the {\em hadronic} level, multiple factorization relations such as
\[
\sigma_{NN}\sigma_{\gamma\gamma}=\left.\sigma_{\gamma N}\right.^2
\]
are expected to hold. 
This implies identical energy dependence for all reactions,
$\sigma\sim s^{\Delta(0)}$, with a universal Pomeron intercept
and independent Pomeron coupling constants for photons and nucleons.
However, this relation is not confirmed by the data.  
In particular, recent HERA measurements of the $\gamma\gamma$ cross 
section have shown a more rapid growth with energy than that seen in 
$pp$ collisions.
In this section show that this feature follows naturally from different 
parton composition, by applying the idea of universal cross sections
at the {\em partonic} level.

At moderate energies, $\sqrt{s}\sim 100$ GeV, the growing part of 
the cross section is small enough so that the simple logarithmic expression 
\begin{equation}
\sigma_{hh'}(s)= \sigma_{hh'} + X_{hh'}\ln(s)
\end{equation}
fits the data quite well. (For definiteness we will use 
values recently fitted by the Particle Data Group 2000 \cite{pdg}.)

Although the second term is small compared to the first,
\[ X_{hh'} << \sigma_{hh'}, \]
one should not assume that the measured $X_{hh'}$ is merely the 
sum of all cross sections involving prompt production. 
Even small partonic cross sections are affected by the screening induced  
by the much larger quasi-elastic processes comprising $\sigma_{hh'}$. 
This is especially clear in the impact parameter plane,  discussed 
by Kopeliovich {\em et al.} \cite{KPPP}. 
Since the nucleon center is nearly black, additional processes 
cannot change the total cross section. 
Therefore, the naive sum of all inelastic processes is always {\em larger} than 
the values present in empirical fits.

In the impact parameter space representation the total cross
section is the integral over the ``blackness'' factor:
\begin{equation}
\sigma_{tot}(s)=2\int d^2b\, \Gamma(b,s).
\label{blackness}
\end{equation}
In the eikonal approximation, blackness is usually represented in the form
\begin{equation} 
\Gamma(b,s)=1-e^{-\chi(b,s)},
\end{equation}
where the quantity in the exponent
is related to the absorption, $\Im A_p(q,s)$, at the ``parton Born level''
(where the subscript $p$ refers to a parton, quark, or gluon).
Specifically, the phase shift function is written
\begin{equation}
\chi(b,s)=\frac{1}{s}\int \frac{d^2q}{(2\pi)}\e^{iq\cdot b}\,\Im A_p(q,s).
\end{equation}
Unitarity is enforced with such a form, in the sense that an increasing
cross section leads to complete blackness, $\Gamma(b,s)\rightarrow 1$.

The next general step is to separate the relatively large and 
$s$-independent quasi-elastic contribution to
$\chi(b,s)$ from the relatively small prompt production part, as
\begin{equation}
\chi(b,s)=\chi_0(b)+\chi_1(b,s).
\end{equation}
After this is done, one expands to first order in $\chi_1(b,s)$ 
and again separates the growing and constant parts of the cross section,
\begin{equation} \label{sigma_tot}
\sigma_{tot}(s)=2\int d^2b \,
\left[ \left( 1-e^{-\chi_0(b)} \right) + \chi_1(b,s)e^{-\chi_0(b)} \right].
\end{equation}
A naive additive formula is recovered for small $\chi_0(b)$, but this is 
in fact not appropriate.  As we will see, corrections to the second
term due to the
$\exp[-\chi_0(b)]$ factor are typically at the 50\% level. This correction is
larger in $pp$ than in $\gamma\gamma$ or $\pi\pi$ collisions, explaining the
apparent differences between the cross sections.

\section{Determining Partonic Content at the Semi-Hard Scale}

Before going into model-dependent studies of shadowing, let us 
first address the partonic composition of different hadrons.

Comparing global fits to hard processes from the literature, one finds
that despite the impressive (and ever-increasing) accuracy of DIS and
Drell-Yan data, there is not yet truly quantitative data on gluons
at the semi-hard scale we need. 
The reason for this is generic, in that gluonic densities
are derived from perturbative DGLAP evolution, which naturally becomes
less accurate as we approach its limits. 
It is even difficult to determine if the density of
semi-hard gluons increases or decreases at small $x$.
Fortunately, for our present purposes we can avoid discussing asymptotically 
large energies, restricting ourselves to sub-TeV domain and including only
partons with Feynman $x>x_{min}= 0.01$. 
The corresponding number of ``relevant partons'' for the nucleon, pion,
and photon are summarized below in Table~\ref{tab_parts}.
The references given in the table are revised GRV parton distributions 
evaluated at next-to-leading order (NLO),
taken at the scale of $Q^2 = 1$ GeV, which are then integrated over interval 
$x = [0.01,1.0]$.

\begin{table}[hbt]
\begin{ruledtabular}
\begin{tabular}{c}
{Proton, with NLO structure functions from Ref.~\cite{GRV_N}}\\
   $N_g$  = 4.10\\
   Valence $N_u$ = 1.70 \\
   Valence $N_d$ = 0.84 \\
   Sea $N_{u+d}$ = 1.16 \\
\colrule
{Pion, with NLO structure functions from Ref.~\cite{GRV_pi}}\\
   $N_g$ = 3.1\\
   Valence $N_{u+\bar{d}}$ = 1.8 \\
   Sea $N_{u+d}$ = 0.48 \\
\colrule
{Photon, with NLO structure functions from Ref.~\cite{GRV_gamma}}\\
   $N_g$ = 1.9 $\alpha$\\
   $N_u  = N_{\bar{u}}$ = 0.87 $\alpha$\\
   $N_d  = N_{\bar{d}}$ = 0.30 $\alpha$\\
\end{tabular}
\end{ruledtabular}
\caption{Partonic content of scattered particles ($\alpha$ is the fine
structure constant).}
\label{tab_parts}
\end{table}

In principle, with more accurate parametrizations,
we might try to test parton additivity by separately extracting, from the data
of the {\em growing} part of hadronic cross section, the contributions
of $qq$, $qg$, and $gg$ to semi-hard processes.
This was attempted, but with the accuracy at hand the 
differences between taking quarks and gluons is negligible.
We are therefore forced to make a model-dependent assumption about their 
relative magnitude. 

Perturbatively, prompt production processes occur during the
collision of two ``wee'' (or Weitzecker-Williams) gluons which accompany 
the large-$x$ partons. 
In the instanton approach this should not necessarily be the case, since
the amplitudes for absorption of various numbers of wee gluons are comparable
and thus there is no suppression by the coupling constant. 
One might try completely resumming the Euclidean Wilson loops, as in
Refs.~\cite{SZ,NSZ}.  We will simply take the lowest order scaling
relation as derived in the next section
and take the effective number of partons to be 
$N_q + 2 N_g$, where $N_q$ and $N_g$ are the numbers of quarks and gluons,
respectively, taken from Table~\ref{tab_parts}.
This leaves us with only one unknown:  
the growing part of the $qq$ cross section.

Combining the parton content with this simple recipe, one obtains the
ratios of cross sections which may be compared to the coefficients of $\ln(s)$
extracted from experiment. 
The results, summarized in Table~\ref{tab_ratios}, are reasonable, but
cannot be taken as precise since shadowing corrections have not been
considered here.

\begin{table}[bth]
\begin{ruledtabular}
\begin{tabular}{ccccc}
& Ratio & Computed & PDG \cite{pdg} &\\
\hline \\
& $ \frac{1}{\alpha} \frac{X_{\gamma N}}{X_{N N}} $    & 0.50 & 0.43 &\\[2mm]
& $ \frac{X_{\pi N}}{X_{N N}} $                        & 0.73 & 0.63 &\\[2mm]
& $ \frac{1}{\alpha} \frac{X_{\gamma N}}{X_{\pi N}}$  & 0.69 & 0.68 &\\[2mm]
& $ \frac{1}{\alpha^2} \frac{X_{\gamma \gamma}}{X_{N N}}$  & 0.25 & 0.16 &
\\[2mm]
\end{tabular}
\end{ruledtabular}
\caption{Cross Section ratios as computed in the text and reported by
the Particle Data Group.}
\label{tab_ratios}
\end{table}

\section{Parton scattering in the Instanton Field}

Multiple parton scattering through an instanton field can be computed either
by evaluating Wilson lines in Euclidean space \cite{NSZ} or using 
perturbative rules in Minkowski space.
Relating these two approaches to all orders is non-trivial.
In this paper we use the former method, analytically continuing 
Wilson lines from Minkowski to Euclidean space as detailed in Ref.~\cite{Meg}
This method maintains diagram-by-diagram correspondence and allows one
to calculate scattering amplitudes involving both quarks and gluons.

The Wilson line in an instanton field $A_\mu^a$ is given by
\begin{equation}
W=P\exp\left(ig\int_{-\infty}^{+\infty} d\tau A^a_\mu(E\tau+r)p^\mu T^a\right),
\end{equation}
where $r$ is distance of Wilson path from the center of instanton and
$T^a$ lies in fundamental representation for quarks and adjoint representation
for gluons. 
Before analytically continuing to Euclidean space let us
introduce small regulatory mass $m$ such that $p^2=m^2$.
Analytic continuation is then performed respecting this condition.
After an obvious change of variables one finds
\begin{equation}
W=P\exp\left(ig\int_{-\infty}^{+\infty} dx A^a_\mu(x+r)
\frac{p^\mu}{E} T^a\right).
\end{equation}
Integrating, for the quark line we have
\begin{equation}
W_{AB}=\cos\left(\alpha\frac{m}{E}\right)\delta_{AB}+
i\frac{E}{m}\sin\left(\alpha\frac{m}{E}\right)n^a\tau^a_{AB},
\end{equation}
where $n^a=\eta^{a\mu\nu}(p_\mu/E)(r_\nu/|r|)$ and
$\alpha=\pi(1-r/\sqrt{r^2+\rho^2})$. 
In the high energy limit, $m/E\rightarrow 0$ leads to
\begin{equation}
W_{AB}=\delta_{AB}+i\alpha n^a\tau^a_{AB}.
\end{equation}
For gluons we similarly have
\begin{eqnarray} \label{Wq}
W_{ab}&=&\cos\left(2\alpha\frac{m}{E}\right)\delta_{ab}
+\frac{E}{m}\sin\left(2\alpha\frac{m}{E}\right)n^c\ve_{abc}\nonumber\\
&&+\left(\frac{E}{m}\right)^2
\left(\cos\left(2\alpha\frac{m}{E}\right)-1\right)n_a n_b\,,
\end{eqnarray}
and in the high energy limit,
\begin{equation}\label{Wg}
W_{ab}=\delta_{ab}+2n^c\ve_{cab}\alpha-2n_a n_b\alpha^2\,.
\end{equation}

Diagrammatical interpretation of these results is straightforward.
In high energy (eikonal) limit only single gluon exchange contributes
to the quark-instanton interaction, whereas single and double exchanges
survive in gluon-instanton interaction.  

Comparing results of Eq.~(\ref{Wq}) to Ref.~\cite{NSZ}, one notices
that the only difference is the change $\sin(\alpha)\rightarrow\alpha$.
The calculational techniques of Ref.~\cite{NSZ} can therefore be used, 
except for a modification of the instanton-induced form factor as 
explained below.

As for gluon scattering, it is a general property of high energy 
cross sections in the instanton field that the number of normal vectors
(the $n_a$) in the cross section corresponds to the moment of relative 
instanton-antiinstanton
rotation and is in turn proportional to $\sqrt{\alpha_s}$ \cite{NSZ}.
Thus, the third term in the r.h.s. of Eq.~(\ref{Wg}) is subleading to 
the order $\alpha_s^2$ (there is no interference between the symmetric 
and antisymmetric terms). 
Such corrections are beyond the present accuracy and will be ignored. 
Consequently, after performing some straightforward
algebra, we find that gluon scattering is governed by a form factor
simply twice that for quarks, meaning that for the purpose
of phenomenology we can consider hadrons as consisting of $N_q+2N_g$ 
``effective quarks''.

\section{Fixing Partonic Cross Sections}

We now proceed with a model-dependent analysis of the problem.
As explained in Section II, we assume that $\chi_1(b,s) \ll 1$
(to be justified with numerics {\em a posteriori}) 
and write the blackness factor of Eq.~(\ref{blackness}) as
\begin{equation}
\Gamma=\Gamma_0+\Gamma_1 \,,
\end{equation}
in which
\begin{equation}
\Gamma_0 = 1 - e^{-\chi_0(b)} 
\, , \quad
\Gamma_1 = \chi_1(b,s) e^{-\chi_0(b)} \,.
\end{equation}
Fourier transforming to a momentum representation, the
rising contribution to the cross section can be written as
\begin{equation}
\chi_1(q,s)_{ij}=\frac{\rho^2}{4}F_{ij}(q)\Delta(q)\ln(s)\,,
\label{rising}
\end{equation}
for two hadrons of types $i$ and $j$ explicitly.
where $F_{ij}(q)$ is the hadronic form factor of H{\"u}fner and Povh
\cite{HP}, and is factorizable as
\begin{equation}
F_{ij}(q)=F_i(q)F_j(q) \,,
\end{equation}
where the single-hadron form factors are of geometric monopole or dipole form
for mesons or baryons, respectively.  
Explicitly, it is the parametrization
\begin{equation}
F_i(q)=\left(1+\frac{q^2R_i^2}{6n_i}\right)^{-n_i},
\label{hpfax}
\end{equation}
with $n=2$ for protons and $n=1$ for mesons. 
The radii, phenomenologically fit to CERN and Fermilab data, 
are $R_p=0.77$ fm and $R_\pi=0.64$ fm.

The function $\Delta(q)$ in Eq.~(\ref{rising}) characterizes the processes 
responsible for prompt production and the growing cross section. 
The overall normalization is chosen so that taking $F(q)=1$ results in
a growing component of $\pi \rho^2 \Delta(0)\ln(s)$, as in Ref.~\cite{NSZ}. 

We now consider the instanton model, in the context of which $\Delta(q)$ 
was calculated in Ref.~\cite{NSZ}, and found to be  
\begin{equation}
\Delta(q)=\kappa\frac{16}{15}\frac{1}{(2\pi)^8}
\int d^2q_1 d^2q_2 \, H(q_1,q_2;q) \,.
\label{delta1}
\end{equation}
The instanton diluteness parameter, $\kappa$, appears linearly and
$H(q_1,q_2;q)$ is the instanton-induced interaction kernel.
The double integral over the kernel may be written as
\begin{equation}
\int \frac{d^2q_1}{(2\pi)^2} 
\frac{d^2q_2}{(2\pi)^2}  \, H(q_1,q_2;q) = 
\left( \int d^2b \, e^{-iq\cdot b} J(b)^2 \right)^2,
\label{kernel}
\end{equation}
where $J(b)$, the Fourier transform of the instanton-induced form
factor, is
\begin{eqnarray}
J(b)&=&\frac{2\pi b}{\rho^2}\int_0^\infty dx \, \frac{1}{\sqrt{x^2+b^2}}
\left(1-\sqrt{\frac{x^2+b^2}{x^2+b^2+\rho^2}}\right)\nonumber\\
&=&\frac{\pi b}{\rho^2}\ln\left(1+\frac{\rho^2}{b^2}\right),\label{aff}
\end{eqnarray}
in which $\rho$ is the average instanton size.

For large $b$, $J(b)\sim 1/b$, and thus the form factor 
is logarithmically divergent.
A finite result can be obtained through phenomenological deformation
of the instanton profile.
We thus replace Eq.~(\ref{aff}) with the deformed instanton form factor,
\begin{equation}
J(b)=\frac{\pi b}{\rho^2}\ln\left(1+\frac{\rho^2}{b^2}\right)\,e^{-cb/\rho} \,.
\label{deformed}
\end{equation}
The constant $c$ is a free parameter, to be fitted to the
total proton-proton cross section.
Combining Eqs.~(\ref{delta1}), (\ref{kernel}), and (\ref{deformed}), we obtain 
\begin{equation}
\Delta(q) = \frac{\kappa}{15\rho^4}
\left( \int d^2b \, {b^2 
\left[\ln\left(1+\frac{\rho^2}{b^2}\right)\right]^2 e^{-2cb/\rho-iq\cdot b}}
\right)^2
\end{equation}

Combining all factors into Eq.~(\ref{rising}),
we have the final phase shift function
\begin{eqnarray}
&&\chi_1(q,s)_{ij} = \nonumber\\
&&\,\,\frac{\rho^2}{4}(N_q+2N_g)_i(N_q+2N_g)_j F_i(q) F_j(q)\Delta(q) \ln(s)\,.
\end{eqnarray}
with constituent numbers $N_\alpha$ taken from Table~\ref{tab_parts}.

We must next consider shadowing corrections to $\chi_1(q,s)$.
Again, following Ref.~\cite{HP}, we use
\begin{equation}\label{chi0}
\chi_0(q)=\frac{\lambda_0}{4\pi}R_1^2R_2^2F_{12}(q) \,,
\end{equation}
where $\lambda_0=0.52$ GeV$^2$ is a universal inverse area for hadronic
collisions.
Finally, we use the standard instanton parameters of
$\kappa=0.01$ and $\rho=0.3$ fm \cite{Shu_82}.

We fix the deformation parameter $c$ by fitting our model's prediction
for the rising part of the $pp$ cross section 
to the experimentally observed one, $X_{pp}=0.174\,{\rm fm^{2}}$
\cite{pdg}. 
This requires $c=0.327$.
We are now able to calculate the rising parts of total cross sections
for other hadrons, and our precitions for $p\pi$, $p\gamma$, and
$\gamma\gamma$ are given in Table~\ref{tab_cs}.
We find reasonable agreement between these numbers and the data,
having fixed only one free parameter, $c$.
\begin{table}[bth]
\begin{ruledtabular}
\begin{tabular}{lrr}
                         & Calculated           & PDG \cite{pdg}\\
\colrule
$X_{p\pi}$               & 0.132                & 0.111\\
$X_{p\gamma}$           & $5.65\times 10^{-4}$ & $5.51\times 10^{-4}$\\
$X_{\gamma\gamma}$       & $1.72\times 10^{-6}$ & $1.45\times 10^{-6}$\\
\end{tabular}
\end{ruledtabular}
\caption{Coefficients $X_{ij} = d\sigma^{tot}_{ij}/d\ln(s)$ in fm$^2$
for different hadronic constituents.}
\label{tab_cs}
\end{table}

\begin{figure}[bt]
\hspace*{-5mm}
\epsfig{file=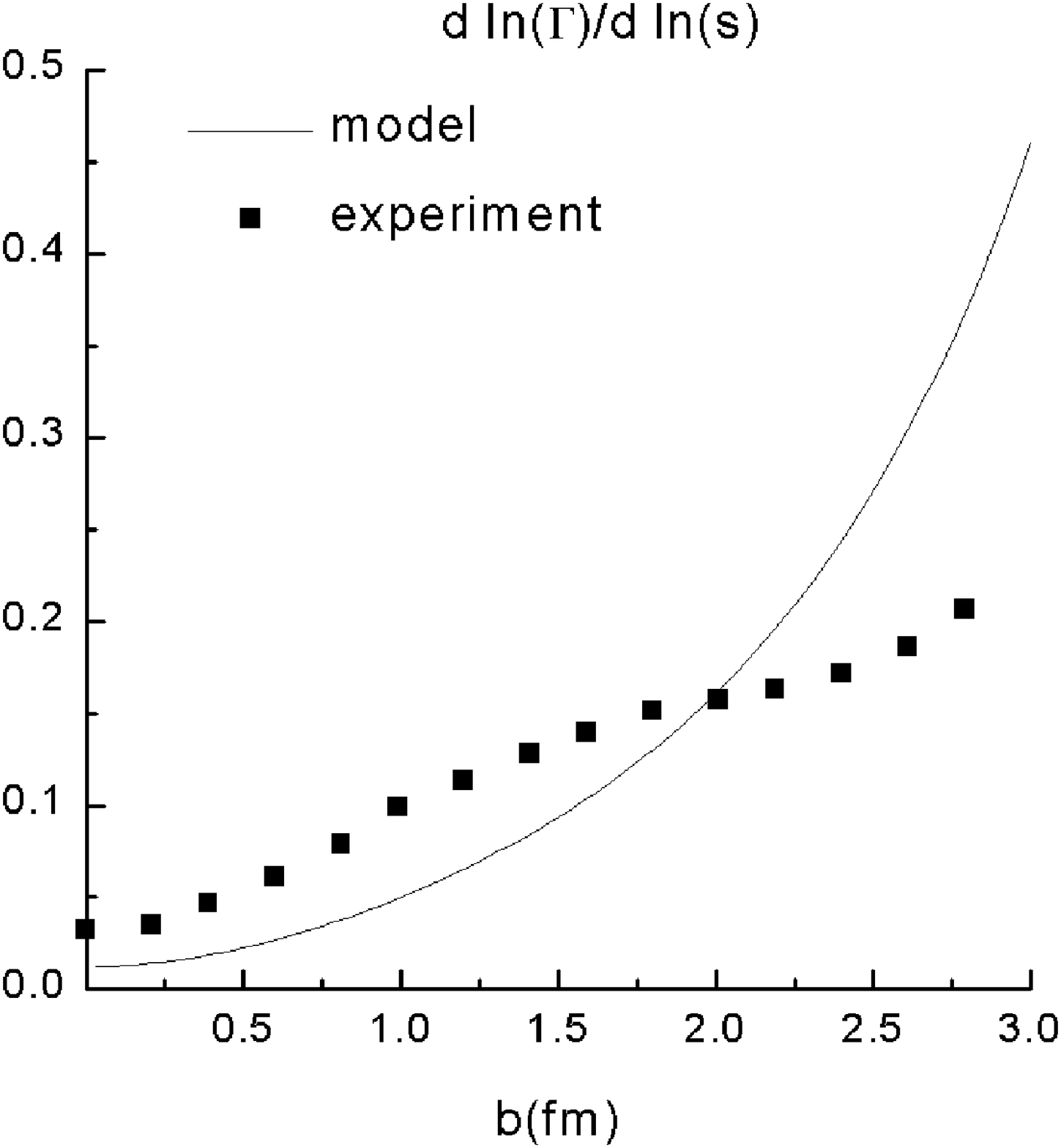, width=80mm}
\caption {
The effective exponent, $\delta(b)$, as a function of the impact 
parameter, $b$.
}\label{fig:dG}
\end{figure}

The $\Gamma(b,s)$ dependence on $b$, which determines the differential 
cross section, is shown in Fig.~\ref{fig:dG}.
Following Ref.~\cite{KPPP}, we have plotted\footnote{
In Ref.~\cite{KPPP} this was defined as $\Delta(b)$; here we use alternative
notation to avoid confusion with our quantity $\Delta(q)$ as defined in
Ref.~\cite{NSZ}.}
\begin{equation}
\delta(b) = \frac{ d\,\ln\Gamma }{d\,\ln s} \,.
\end{equation}
The experimental points are a parametrization fit done by Kopeliovich {\em et
al}. with ISR \cite{isr} and CERN UA4 \cite{sps} data.
While our model captures the overall systematics of the curve extracted
from experiment, it is clearly lacking at large $b$, suggesting that
we have overestimated the instanton tail contribution.

\section{Heavy Ion Collisions}

We may now extend our analysis to estimate the number of sphaleron-type
clusters produced from excited instantons in heavy ion collisions. 
This issue has already been
discussed by one of us \cite{Shu_AA}, and we now return to it
with more definite knowledge of the parameters involved. 

For symmetric, central $AA$ collisions of two nuclei we use the
simplest model, one of two spheres with homogeneously distributed partons.
The total parton number is $AN_q$, with  $N_q\approx 12$ being the number of
``effective quarks'' (quarks number plus twice gluons number)
per nucleon\footnote{Of course, the clustering of partons
into ``constituent quarks'' and nucleons increases the number of
collisions, but we will ignore such correlations for now.}.

The total number of $qq$ collisions in this case is easily obtained from 
the following geometric integral:
\begin{eqnarray}
N_{coll} &=& 8\pi\sigma_{qq} n_q^2\int_0^R dr_t r_t \left(R^2-rt^2\right)
\nonumber\\
&=& 3^{4/3} 2^{-5/3} \pi \sigma_{qq} n_q^2 \left({A N_q \over \pi n_q}
\right)^{4/3},
\end{eqnarray}
where the quark density is determined by the nuclear density to be
$n_q = N_q\times 0.16$ fm$^{-3}$.

With $A=197$ (gold) and the value for the quark-quark cross section
extracted above, $\sigma_{qq}=1.69\times 10^{-3}$ fm$^2$, we have the
following production rate per unit rapidity of sphaleron-like clusters:
\begin{equation}
\frac{dN_{coll}}{dy}  \approx 76.5 \,,
\end{equation}
a number somewhat smaller than estimated in Ref.~\cite{Shu_AA}.

Each cluster will in turn decay into a number of quarks and gluons.
Simply scaling of the couplings from
the studies of sphaleron decay in electroweak theory leads to about 3.5
gluons per cluster, with 0-6 quarks 
(up to a complete set of light quark-antiquark pairs,
$\bar u u \bar d d \bar s s$). 
As an average we tentatively take 3.5 gluons and 2.5 quarks,
the latter obtained by applying a factor of one half for the suppression of 
strange quarks and another one half to account for the possibly change in
Chern-Simons number.
This yields an average of six partons per cluster, or in central $AuAu$ 
collisions at RHIC about $76.5\times 6=460$ partons per rapidity from 
sphaleron production.
This is roughly {\em one half} the maximal possible value, 
$dN_{partons}/dy\sim dN_{hadrons}/dy\sim
1000$, inferred experimentally from the final entropy limitations.

This result is in good agreement with phenomenological studies of the 
energy and impact-parameter dependence of multiplicity \cite{KN}, which have
deduced that the contribution to multiplicity which scales as
the number of parton collisions generates about half of the total,
when calculated from the standard Glauber model and
using the experimental nuclear density distribution for a gold nucleus.
In this picture, the $\sim 500$ hadrons per unit rapidity are then a result
of prompt production from QCD sphalerons.

The competing mini-jet picture, in which the products come from 
hadronization of two mini-jets, can also explain this number. 
However, in the minijet scenario this hadronization occurs later
and one must confront problems such as the origin of strong collective effects,
jet quenching, and the pronounced ellipticity at large $p_t$ observed at RHIC.

\section{Summary and Discussion}

The objective of this paper was to evaluate the magnitude
of the parton-parton cross sections which lead to prompt production
and contribute to the growth of hadronic cross sections involving protons,
pions, and photons.
We have demonstrated that, at the semi-hard scale of $Q^2 \sim 1$ GeV,
it is vital to accurately estimate the partonic content of each 
scattered particle.
Furthermore, it was shown that a significant part of these effects are
hidden by screening or quasi-elastic color exchange processes, {\em i.e.}
the constant part of the cross sections.

The main assumption was that
partons act additively (or, more precisely, multiplicatively), thereby 
ignoring transverse correlations which might reduce the cross sections.
The nonperturbative dynamics were computed in the instanton model, taking
formulae derived in Ref.~\cite{NSZ}.

Our main result is a surprisingly small prompt production component
in parton-parton cross sections, reported in the Table~\ref{tab_cs}, 
on the order of
\begin{equation} 
\sigma_{qq} \sim 10^{-3}\, {\rm fm}^2\,.
\end{equation}
Naive geometric cross sections are 300 times larger, and thus an explanation
of this much smaller number is necessary.

In terms of the instanton picture, the diluteness of the instanton
ensemble, $n\rho^4\sim 10^{-2}$, is in fact insufficiently small.
An additional suppression thus seems to be needed. 
Following Ref.~\cite{NSZ}, rather than changing the phenomenologically 
sound parameters of the instanton ensemble, we instead turned to 
instanton tail suppression through an
{\em ad hoc} exponential factor of $exp\left(-M\vert x\vert\right)$. 
The results imply a rather large $M$ of about 500 MeV\footnote{
Incidentally, this is close to the mass suggested on the 
basis of the mutual instanton repulsion \cite{DP}.}.

We have shown that with this small cross section one can reasonably describe 
hadronic data -- the energy dependence as
a function of impact parameter and the growing parts of $NN$, $\pi N$, $\gamma
N$, and $\gamma\gamma$ cross sections -- and roughly estimate
the amount of entropy produced in high energy $AuAu$ collisions at RHIC.
In the future we plan to make this scenario much more quantitative by 
not only calculating the average parton numbers from sphaleron decay, but
also their momentum spectra, quark/gluon ratios, and more.

\begin{acknowledgments}
We thank I. Zahed for useful discussions.
This work was partially supported by US-DOE grants DE-FG02-88ER40388
and DE-FG03-97ER4014.
\end{acknowledgments}

\end{document}